\documentclass[aps,pra,twocolumn,superscriptaddress,10pt]{revtex4-2}
\usepackage{amsmath,amssymb,graphicx,bm}
\usepackage[dvipsnames]{xcolor}

\begin{document}

\title{Experimental constraints on gauge hidden-variable models in Bell tests}

\author{Li-Yu Shen}
\thanks{These authors contributed equally to this work.}
\affiliation{Institute for Quantum Studies, Chapman University, Orange, California 92866, USA}
\affiliation{Nexus for Quantum Technologies, University of Ottawa, Ottawa, Ontario K1N 6N5, Canada}

\author{Lukas Scarfe}
\thanks{These authors contributed equally to this work.}
\affiliation{Institute for Quantum Studies, Chapman University, Orange, California 92866, USA}
\affiliation{Nexus for Quantum Technologies, University of Ottawa, Ottawa, Ontario K1N 6N5, Canada}

\author{Carrie Rodriguez}
\affiliation{Institute for Quantum Studies, Chapman University, Orange, California 92866, USA}

\author{Alessio D'Errico}
\affiliation{Institute for Quantum Studies, Chapman University, Orange, California 92866, USA}
\affiliation{Nexus for Quantum Technologies, University of Ottawa, Ottawa, Ontario K1N 6N5, Canada}

\author{Andrea Aiello}
\affiliation{Max Planck Institute for the Science of Light, Staudtstrasse 2, 91058 Erlangen, Germany}

\author{Ebrahim Karimi}
\affiliation{Institute for Quantum Studies, Chapman University, Orange, California 92866, USA}
\affiliation{Nexus for Quantum Technologies, University of Ottawa, Ottawa, Ontario K1N 6N5, Canada}
\affiliation{Max Planck Institute for the Science of Light, Staudtstrasse 2, 91058 Erlangen, Germany}

\begin{abstract}
Quantum mechanics assigns identical statistics to measurement procedures that realize the same final effect, irrespective of the transformations used to reach it. We test this operational equivalence with polarization-entangled photons by implementing one of Bob's analyzer settings either directly or through closed wave-plate trajectories $C_\gamma$ calibrated to produce the same measurement operator. Motivated by ontological descriptions in which a hidden frame is transported between contexts, we formulate a local path-memory model and derive the mixed-context bound $|S_\gamma^{\rm mix}|\leq 2+2p_\gamma$. The quantity $p_\gamma$ is a coupling-dependent counterfactual mismatch probability: it bounds a contextual Bell expression, but neither generates nor predicts the quantum Bell correlation. We therefore test path memory through the observable difference between direct and looped implementations, using a no-signaling phenomenological response in which a hidden holonomy $\delta_\gamma$ modifies one correlation term in $S_\gamma^{\rm mix}$. The direct reference violates the Bell inequality, while the looped mixed-context curves remain indistinguishable from it within the experimental precision. Our results support the quantum equivalence of calibrated measurement implementations and constrain path-sensitive hidden-variable extensions that predict an observable memory of the optical history.
\end{abstract}

\maketitle

\noindent\emph{Introduction.}
Bell's theorem turned the tension between entanglement and local realism into an experimental question~\cite{Einstein1935,Schrodinger1935,Bell1964}. In the Clauser-Horne-Shimony-Holt (CHSH) scenario, Alice chooses between settings $a$ and $a'$, Bob between $b$ and $b'$, and the signed combination $S=E(a,b)+E(a,b')+E(a',b)-E(a',b')$ is formed~\cite{CHSH1969}. Any Bell-local model with setting-independent hidden variables obeys $|S|\leq2$, whereas quantum mechanics allows $2\sqrt2$. Loophole-free experiments have established violations of the Bell-local bound~\cite{Hensen2015,Giustina2015,Shalm2015}; the question is no longer whether entangled correlations can be explained by a Bell-local model.

Here, we address a different assumption, usually left implicit: what physically constitutes a measurement setting? Quantum theory identifies the setting with its measurement effects. In an optical experiment, however, the same effect can be realized through many sequences of wave plates and polarizers. Quantum mechanics treats these implementations as operationally equivalent. An ontological extension need not do so if an internal degree of freedom retains a memory of the path used to reach the final analyzer.

We formulate this possibility using a gauge hidden-variable (GHV) description of polarization. A pure polarization state defines a point on the Poincar\'e sphere, and an oriented frame is attached to the corresponding tangent plane.
A connection transports the hidden frame along the trajectory traced by the optical transformations. After a closed path, the quantum measurement can return to its initial form while the hidden frame acquires a holonomy set by the enclosed solid angle~\cite{Yang1954,Berry1984,Nakahara2003}.

This geometry does not evade Bell's theorem. If the complete response remains Bell factorisable, it obeys the Bell bound regardless of the topology of its hidden-variable space. We use the gauge construction for a narrower purpose: to identify a possible path-dependent correction to an already observed entangled correlation. We compare direct and looped implementations of Bob's analyzer, calibrated to realize the same final effect, and search for a structured deformation of a mixed-context witness. Such a deformation is distinct from a loss of visibility or an ordinary wave-plate calibration error.

\begin{figure}
    \includegraphics[width=\linewidth]{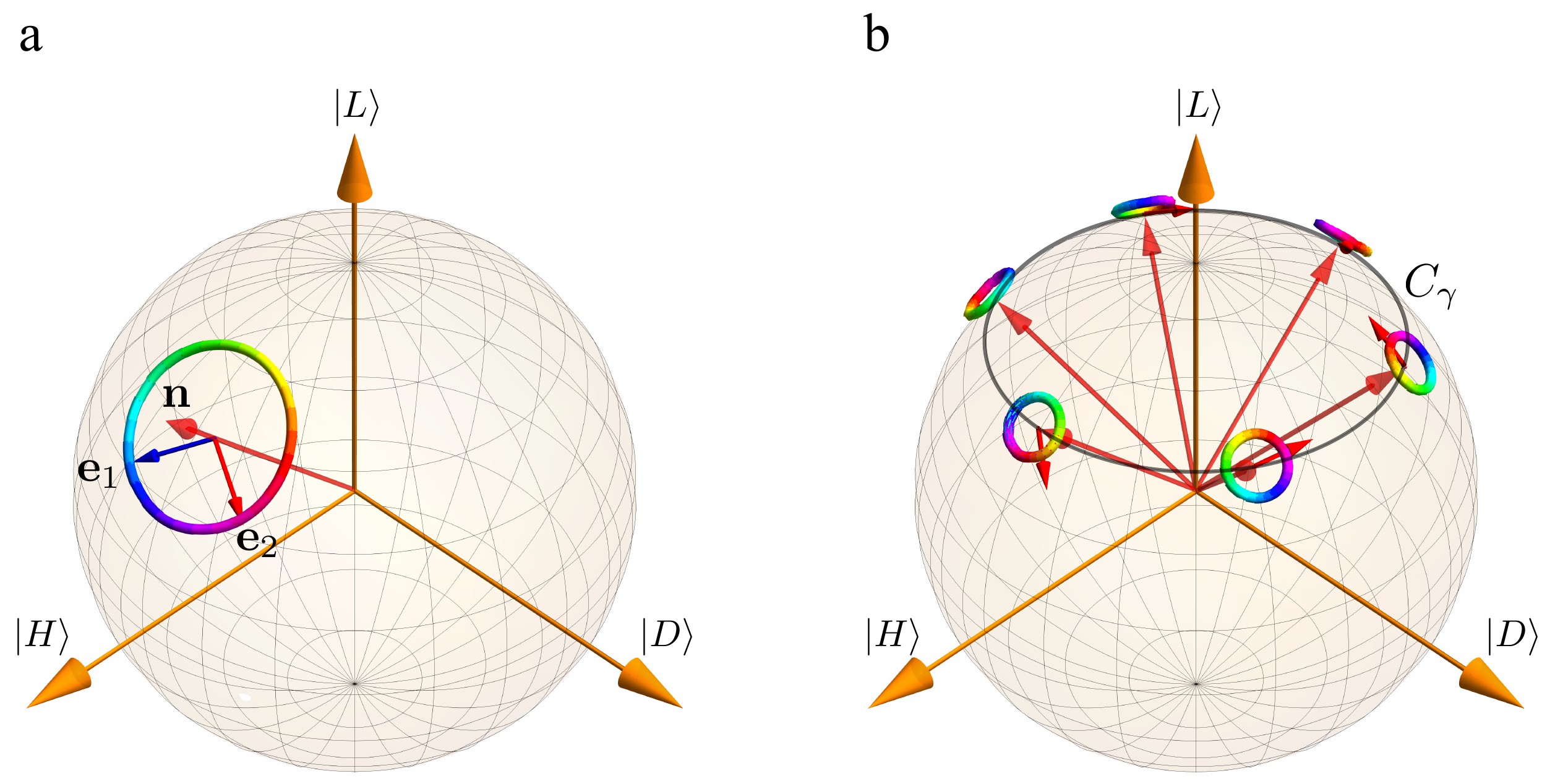}
    \caption{\textbf{Geometric transport of a hidden tangent frame.} \textbf{a}, A pure polarization state $\mathbf n$ is represented by a point on the Poincar\'e sphere $S^2$. The orthonormal basis $\{\mathbf e_1,\mathbf e_2\}$ spans the local tangent plane, and the colored circle denotes the $U(1)$ fiber of hidden orientations, $\mathbf v_\theta=\cos\theta\,\mathbf e_1+\sin\theta\,\mathbf e_2$. \textbf{b}, Bob's polarization follows a closed path $\mathcal C_\gamma$. The hidden frame is transported between successive tangent planes and, after the polarization returns to its initial state, may acquire the holonomy $\delta_\gamma=\kappa\,\Omega_\gamma$, where $\Omega_\gamma$ is the enclosed solid angle. Operationally equivalent paths can therefore retain different histories in the GHV description.}
    \label{fig:fig1}
\end{figure}

\noindent\emph{Gauge-covariant local response.}
In both classical and quantum contexts, a polarizer can be characterized by a unit vector $\mathbf{n} $ on the Poincar\'e sphere. In quantum mechanics, the action of such a polarizer can be represented by the binary projection operator $\Pi_{t|\mathbf{n}} = \frac{1}{2}(I + t\, \mathbf{n} \cdot \bm{\sigma} ), \; t = \pm 1 $, 
where $I$ is the identity matrix, $\bm{\sigma} = (\sigma_x, \sigma_y, \sigma_z) $ are the Pauli matrices, and $\pm \mathbf{n}$ describe two orthogonal states of polarization. If we put a sequence, say $\Gamma$, of lossless waveplates along the path of light before entering the polarizer, these plates can be represented by a unitary operator $U[\Gamma]$, such that $\Pi_{t|\mathbf{n}} \to \Pi^{(\Gamma)}_{t|\mathbf{n}} = U^\dagger[\Gamma] \Pi_{t|\mathbf{n}} U[\Gamma]$. In quantum mechanics, two measurement procedures are considered \emph{operationally equivalent} if they produce the same measurement operators, and thus, yield identical probabilities for all outcomes. Thus, two waveplates' sequences $\Gamma_0$ and $\Gamma_\gamma$ are operationally equivalent when $\Pi_{t|\mathbf{n} }^{(\Gamma_0)}=\Pi_{t|\mathbf{n}}^{(\Gamma_\gamma)}$. 

Given the Poincar\'e sphere $S^2$, every Stokes vector $\mathbf{n} \in S^2$ has a tangent space $T_{\mathbf{n}}S^2$. The unit tangent bundle $UTS^2$ is the collection of all tangent vectors of length $1$ at all points.
Here, the GHV ontic space is taken to be the unit tangent bundle $UTS^2 = \{(\mathbf{n},\mathbf{v})\mid \mathbf{n} \in S^2,\,\mathbf{v}\in T_{\mathbf{n}}S^2,\,\|\mathbf{v}\|=1\}$. We use the symbol $\pi$ to represent the natural projection map $\pi : UTS^2 \rightarrow S^2$, associated with this bundle, such that $\pi (\mathbf{n},\mathbf{v}) = \mathbf{n}$.
Let the unit vector $\mathbf{n}(s) \in S^2$ trace a closed curve $C_\gamma$ on the unit sphere, where $0 \leq s \leq 1$ is a continuous parameter such that $\mathbf{n}(0) = \mathbf{n}(1)$ (closure condition). To each value of $s$ is associated a tangent space $T_{\mathbf{n}(s)}S^2$. For such a space we introduce the orthonormal basis $ \{ \mathbf{e}_1(s), \mathbf{e}_2(s), \mathbf{n}(s) \}$. Let $\mathbf{v}_\theta(s)$ be a tangent vector field along $C_\gamma$, defined by $\mathbf{v}_\theta(s)=\cos\theta\,\mathbf{e}_1(s)+\sin\theta \,\mathbf{e}_2(s)$, where $\theta \in [0, 2 \pi)$. We call $\mathbf{v}_\theta(s)$ the \emph{hidden orientation} field.
The vector $\mathbf v_\theta(s)$ is precisely a choice of lift, parameterized by $\theta$, of the curve $\mathbf{n}(s)\subset S^2$ into $UTS^2$: at each $s$ it specifies not just the point $\mathbf{n}(s)$ but also a direction in the tangent plane, i.e.,\ a point of the fiber $\pi^{-1}(\mathbf{n}(s))\cong S^1$. Thus the pair $\left(\mathbf{n}(s),\,\mathbf v_\theta(s)\right)\in UTS^2$ contains all the information of the polarization state characterized by the Stokes vector $\mathbf{n}(s)$,  except its overall sign. 

When $s$ varies between $0$ and $1$, the basis vectors $ \mathbf{e}_1(s) $ and  $\mathbf{e}_2(s)$  undergo a parallel transport along $C_\gamma$. By definition, $\mathbf{v}_\theta(s)=\cos\theta\,\mathbf{e}_1(s)+\sin\theta \,\mathbf{e}_2(s)$, so that $\theta$ is \textbf{invariant} under parallel transport. However, the hidden orientation field rotates by the angle $\Omega_\gamma$, as $\mathbf v_\theta(1)=\cos(\theta+\Omega_\gamma)\,\vec e_1(0)+\sin(\theta+\Omega_\gamma)\,\vec e_2(0)$,  where $\Omega_\gamma$ is the  solid angle subtended by the closed curve $C_\gamma$. If $\bm{p}(s) \in \mathbb{C}^2$ is the Jones vector generating the Stokes vector $\mathbf{n}(s)$ via the Pauli-matrix map $n_i(s) = \bigl(\bm{p}(s), \sigma_i \bm{p}(s) \bigr)$, $(i=1,2,3)$, here, $(\cdot,\cdot)$ is the scalar product. Then under parallel transport $\mathbf{n}(1) = \mathbf{n}(0)$, and $\bm{p}(1) = e^{- i \Omega_\gamma/2}\bm{p}(0)$. The Stokes vector $\mathbf{n}(s)$ and the Jones vector $\bm{p}(s)$ belong to different vector spaces and carry different amounts of information, with $\bm{p}(s)$ giving the more complete description of polarization.

In this context, a polarisation analyser-- whether Alice’s or Bob’s --is fully characterized, apart from a global sign, by the pair $x=(\mathbf{n},\Gamma)$, where $+\mathbf{n}$ and $-\mathbf{n}$ are the Stokes vectors describing the polarization transmitted and reflected by the polarizer, respectively, and the waveplates sequence $\Gamma$ determines the closed curve $C_\gamma$. 

Suppose the light source emits a pair of photons in the ``hidden state'' $\eta(\theta)$, and $\rho(\theta) \text{d} \theta$ is the probability measure assigned to this state.
Let $p_B(t | x_B, \theta)$ be the probability that for a given photon pair emitted in the state  $\eta(\theta)$, Bob's detector will fire $(t = +1)$, or will not fire $(t = -1)$, where $x_B$ specifies the analizer setting, e.g., the polarizer orientation. A simple local detector model suggests taking $p_B(t | x_B, \theta)=\bigl( 1+ t \, \mathbf{n}_B \cdot \mathbf{V}_\theta  \bigr)/2$, where $\mathbf{V}_\theta  \in \mathbb{R}^3$ is a unit vector defined by $\mathbf{V}_\theta = \bigl[\mathbf{v}_\theta (0) \cdot \mathbf{e}_1(0) \bigr] \mathbf{e}_x + \bigl[ \mathbf{v}_\theta (0) \cdot \mathbf{e}_2(0)\bigr] \mathbf{e}_y $. For a linear analyzer at angle $\beta$, we have $\mathbf{n}_B = \cos(2 \beta) \mathbf{e}_x +  \sin(2 \beta) \mathbf{e}_y$, and since $\mathbf{v}_\theta (0) = \cos \theta \, \mathbf{e}_1(0) + \sin \theta \, \mathbf{e}_2(0)$, then
\begin{equation}
    p_B(\pm|\beta,\theta) = \left\{
                              \begin{array}{ll}
                                \cos^2\left(\beta-\frac{\theta}{2}\right), & \text{for } + \\[3pt]
                                \sin^2\left(\beta-\frac{\theta}{2}\right), & \text{for } - 
                              \end{array}
                            \right.
    \label{eq:ghv-hidden-malus}
\end{equation}
When a waveplates sequence $\Gamma$ is in place, $\mathbf{n}_B$ remains unchanged, and $ \mathbf{V}_\theta \to \mathbf{V}_{\theta + \Omega_\gamma}$, so that $p_B(\pm|\beta,\theta) \to p_B(\pm|\beta,\theta_\gamma)$, where we have defined  $\theta_\gamma = \theta + \Omega_\gamma$. These results also apply, mutatis mutandis, to $p_A(\pm|x_A,\theta) = p_A(\pm|\alpha,\theta)$, where $\alpha$ is Alice's linear analyzer orientation. Bell locality is imposed by the requirement that the complete model be Bell factorisable, that is
\begin{equation}
    p_{AB}(s,t|x_A,x_B, \theta) =    p_A(s|x_A,\theta)p_B(t|x_B,\theta),
    \label{eq:ghv-bell-local-joint}
\end{equation}
with $s,t = \pm 1$.

The observed probabilities $P_A(s|x_A)$, $P_B(t|x_B)$, and $ P_{AB}(s,t|x_A,x_B)$, $(s,t = \pm 1)$, are obtained by averaging $p_A(s|x_A,\theta)$, $p_B(t|x_B,\theta)$, and $p_{AB}(s,t|x_A,x_B,\theta)$, respectively, over the probability distribution $\rho(\theta)$ for the
hidden variables. Choosing a uniform fiber distribution $\rho(\theta)=1/(2\pi)$, we find  $P_A(s|\alpha) = 1/2 = P_B(t|\beta)$, and $ P_{AB}(s,t|\alpha, \beta) = \frac{1}{4} \bigl\{ 1 + s \, t \, \frac{1}{2} \cos\left[ 2(\alpha-\beta) \right] \bigr\}$. By construction $P_A(s|\alpha)  = \sum_{t = \pm 1} P_{AB}(s,t|\alpha, \beta) = 1/2$. 
The shift $\theta\mapsto\theta+\Omega_\gamma$ preserves the observed probabilities. A hidden frame can therefore retain a local path memory without changing Alice and Bob's single-photon count rates.

To connect these results with the standard quantum-mechanical prediction, let us define the dichotomic random variables $A$ and $B$, which take the values $\pm 1$ with the joint probability $ P_{AB}(s,t|\alpha, \beta)$. By definition $\mathbb{E}[A] = 0 = \mathbb{E}[B]$, and $\mathbb{E}[A^2] = 1 = \mathbb{E}[B^2]$. The correlation $\mathbb{E}[A B]$ is thus given by,
\begin{equation}
    \mathbb{E}[A B] = \frac{1}{2} \cos[2(\alpha - \beta)].
    \label{eq:ghv-local-baseline}
\end{equation}
The angular dependence is in agreement with the quantum-mechanical prediction, but the visibility is one half and the largest sinusoidal CHSH value is $\sqrt{2}$. More generally, no Bell-factorisable completion can exceed $|\mathcal{S}|=2$. The tangent-frame model supplies local Malus statistics and a geometric memory; it does not derive the unit-visibility quantum correlation.\\

\noindent\emph{Counterfactual mismatch bound.}
Path memory allows the response associated with the nominal setting ${\mathbf{n}_B'}$ to depend on its implementation. For the direct $(\Gamma_0)$ and looped $(\Gamma_\gamma)$ sequences, define the two dichotomic random variables $b', b'_\gamma \in \{-1,1\}$, as
\begin{align}
    b'&=b({\mathbf{n}_B'},\Gamma_0,\theta) \simeq p_B(\pm | \beta', \theta),\nonumber\\
    b'_\gamma &=b({\mathbf{n}_B'},\Gamma_\gamma,\theta) \simeq p_B(\pm | \beta', \theta_\gamma) ,
    \label{eq:ghv-two-implementations}
\end{align}
where $\theta_\gamma = \theta + \Omega_\gamma$, and here and hereafter we use the same notation as in Eq. \eqref{eq:ghv-hidden-malus}, but now the probabilities $0 \leq p_B(\pm|\cdots) \leq 1$, with $p_B(+|\cdots) + p_B(-|\cdots) =1$,   can be arbitrary.  
For a complete hidden state, including any random seed used to make the response deterministic, consider the mixed-context polynomial
\begin{equation}
    M_\gamma(\theta) = ab+ab'+a'b-a'b'_\gamma,
    \label{eq:ghv-contextual-chsh}
\end{equation}
where the dichotomic random variables $a = a({\mathbf{n}_A},\Gamma_0,\theta)$$ \sim p_A(\pm | \alpha, \theta)$, $a' = a({\mathbf{n}_A'},\Gamma_0,\theta) $$ \sim p_A(\pm | \alpha', \theta)$, and $b = b({\mathbf{n}_B},\Gamma_0,\theta) $$ \sim p_B(\pm | \beta, \theta)$, also take the values $-1$ and $1$.
Adding and subtracting $a'b'$ in the equation above gives
\begin{align}
     M_\gamma(\theta)=\left(ab+ab'+a'b-a'b'\right) +a'(b'-b'_\gamma).
    \label{eq:ghv-contextual-split}
\end{align}
The term in parentheses is a standard deterministic CHSH polynomial and has magnitude $\leq 2$. Similarly, $-2 \leq a'(b'-b'_\gamma) \leq 2$, so that
\begin{equation}
    |M_\gamma(\theta)| \leq  \begin{cases} 
4, & b' \neq b_\gamma', \\ 
2, & b' = b_\gamma'. 
\end{cases}
    \label{eq:ghv-pointwise-bound}
\end{equation}
Defining 
\begin{align}
p_\gamma(\theta)  = \text{Prob} \bigl(b'\neq b'_\gamma \bigr) = \frac{ 1 - \mathbb{E}[b'] \, \mathbb{E}[b_\gamma']}{2} ,
    \label{eq:p-gamma}
\end{align}
and $S_\gamma^{\rm mix}=\langle M_\gamma (\theta)\rangle_{\rho(\theta)}$, yields
\begin{equation}
    |S_\gamma^{\rm mix}|
    \leq2+2 \, p_\gamma.
    \label{eq:ghv-contextual-bound}
\end{equation}
Equation~\eqref{eq:ghv-contextual-bound} is not the standard two-setting CHSH inequality. The two occurrences of ${\mathbf{n}_B'}$ are different complete settings, distinguished by their physical implementations. Moreover, $p_\gamma$ only records how often the two counterfactual outcomes differ. It does not determine the sign of the correction $a'(b'-b'_\gamma)$ or its correlation with the remaining terms. The condition $p_\gamma\geq\sqrt2-1$ is therefore necessary for the upper bound to accommodate $2\sqrt2$, but it is not sufficient to produce that value. At $\gamma=0$, one has $p_0=0$, and the local construction still obeys the Bell bound; it cannot account for the observed reference violation.

Using the local response model Eqs.~\eqref{eq:ghv-hidden-malus} and \eqref{eq:ghv-bell-local-joint}, we obtain the probability distributions of $a, a', b, b'$ and $b'_\gamma$, we can calculate $\langle p_\gamma \rangle_{\rho(\theta)}$ from \eqref{eq:p-gamma}, to obtain
\begin{align}\label{eq:p-gamma-bis}
P_\gamma^{\text{loc}}  = \frac{1}{2 \pi} \int_0^{2 \pi}  p_\gamma(\theta) \text{d} \theta = \frac{1}{4} + \frac{1}{2} \sin^2 \left( \frac{\Omega_\gamma}{2} \right) , 
\end{align}
where $\rho(\theta) = 1/(2 \pi)$ has been used.
This is a property of the chosen local coupling, not a general observable prediction. The Malus marginal alone does not determine $p_\gamma$, and Eq.~\eqref{eq:p-gamma-bis} should not be identified with the non-factorisable response introduced below. 
\begin{figure*}
    \includegraphics[width=0.8\textwidth]{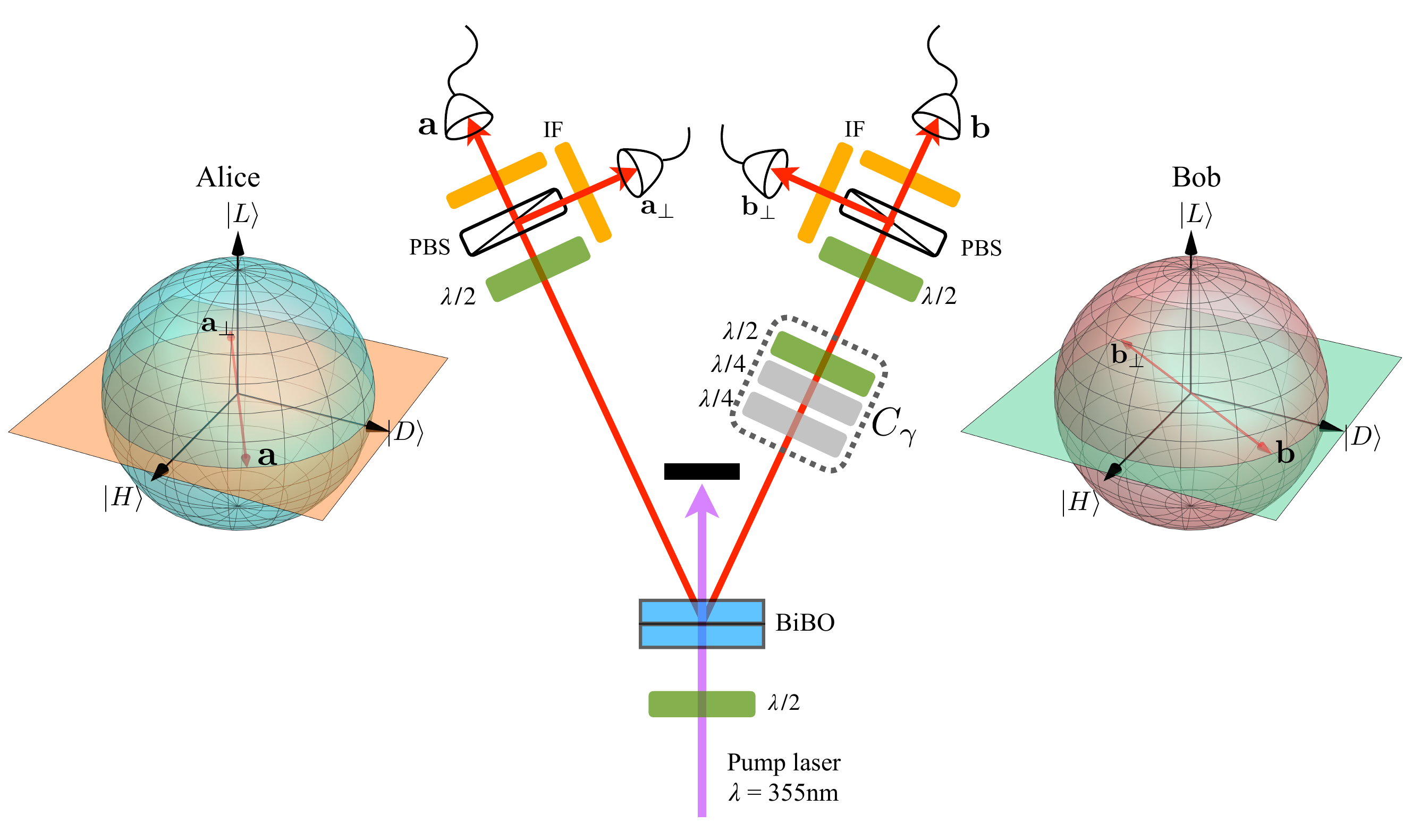}
    \caption{\textbf{Experimental implementation.} A $355~\mathrm{nm}$ laser pumps two orthogonally oriented BiBO crystals to generate polarization-entangled photon pairs by spontaneous parametric down-conversion. Alice performs direct linear-polarization measurements. Bob realizes the same nominal analyzer either directly or through a QWP--QWP--HWP sequence that traces a closed trajectory on the Poincar\'e sphere. The direct and looped sequences are calibrated to produce the same final measurement effects. Quantum mechanics, therefore, predicts identical coincidence statistics; an implementation-dependent difference would signal additional path memory.}
    \label{fig:experiment}
\end{figure*}

\noindent\emph{Operational response tested experimentally.}
The experiment does not use $p_\gamma$ to generate the baseline Bell violation. Instead, it asks whether the observed entangled correlation acquires an additional dependence on the implementation path. The source produces a state close to
\begin{equation}
    |\Phi^+\rangle
    =\frac{1}{\sqrt2}\left(|HH\rangle+|VV\rangle\right),
    \label{eq:ghv-phi-minus}
\end{equation}
for which the direct correlation is written $E_0(\alpha,\beta)=V\cos[2(\alpha-\beta)]$, where $0\leq V\leq 1$ is the correlation visibility. To parameterize a possible path-dependent displacement while preserving unbiased marginals, we use
\begin{equation}
    P_\gamma(s,t|\alpha,\beta)
    =\frac14[1+stE_\gamma(\alpha,\beta)],
    \label{eq:ghv-operational-joint}
\end{equation}
with
\begin{equation}
    E_\gamma(\alpha,\beta)
    =V\cos[2(\alpha-\beta)-\delta_\gamma],
    \label{eq:ghv-loop-correlation}
\end{equation}
where
\begin{equation}
    \delta_\gamma=\kappa\,\Omega_\gamma
    \quad(\mathrm{mod}\ 2\pi),
    \label{eq:ghv-holonomy-coupling}
\end{equation}
and $\kappa$ is a phenomenological response coefficient. The value $\kappa=0$ gives path independence and $\kappa=1$ corresponds, up to orientation, to Levi-Civita transport of a tangent vector on the unit sphere. Other values require additional ontological structure and do not follow from the Pancharatnam-Berry phase alone.

Equation~\eqref{eq:ghv-operational-joint} is no-signaling but, for Bell-violating correlations, not Bell factorisable. It is a phenomenological joint response used to expose a measurable signature of hidden-frame transport; it is not derived by saturating Eq.~\eqref{eq:ghv-contextual-bound}. In particular, $P_\gamma^{\rm loc}$ does not determine the visibility or phase of Eq.~\eqref{eq:ghv-loop-correlation}.

The correlation difference $\Delta E_\gamma=E_\gamma-E_0$ is
\begin{equation}
    \Delta E_\gamma
    =2V\sin\left(\frac{\delta_\gamma}{2}\right)
    \sin\left[2(\alpha-\beta)-\frac{\delta_\gamma}{2}\right].
    \label{eq:ghv-correlation-shift}
\end{equation}
For $|\delta_\gamma|\ll1$, this becomes $\Delta E_\gamma\simeq V\delta_\gamma\sin[2(\alpha-\beta)]$. The first-order sensitivity is largest near a zero crossing of the correlation fringe and vanishes at its extrema. A nonzero $\Delta E_\gamma$ would establish operational path dependence, but a null result does not place a model-independent upper bound on the counterfactual quantity $p_\gamma$.

\begin{figure*}
    \includegraphics[width=0.9\textwidth]{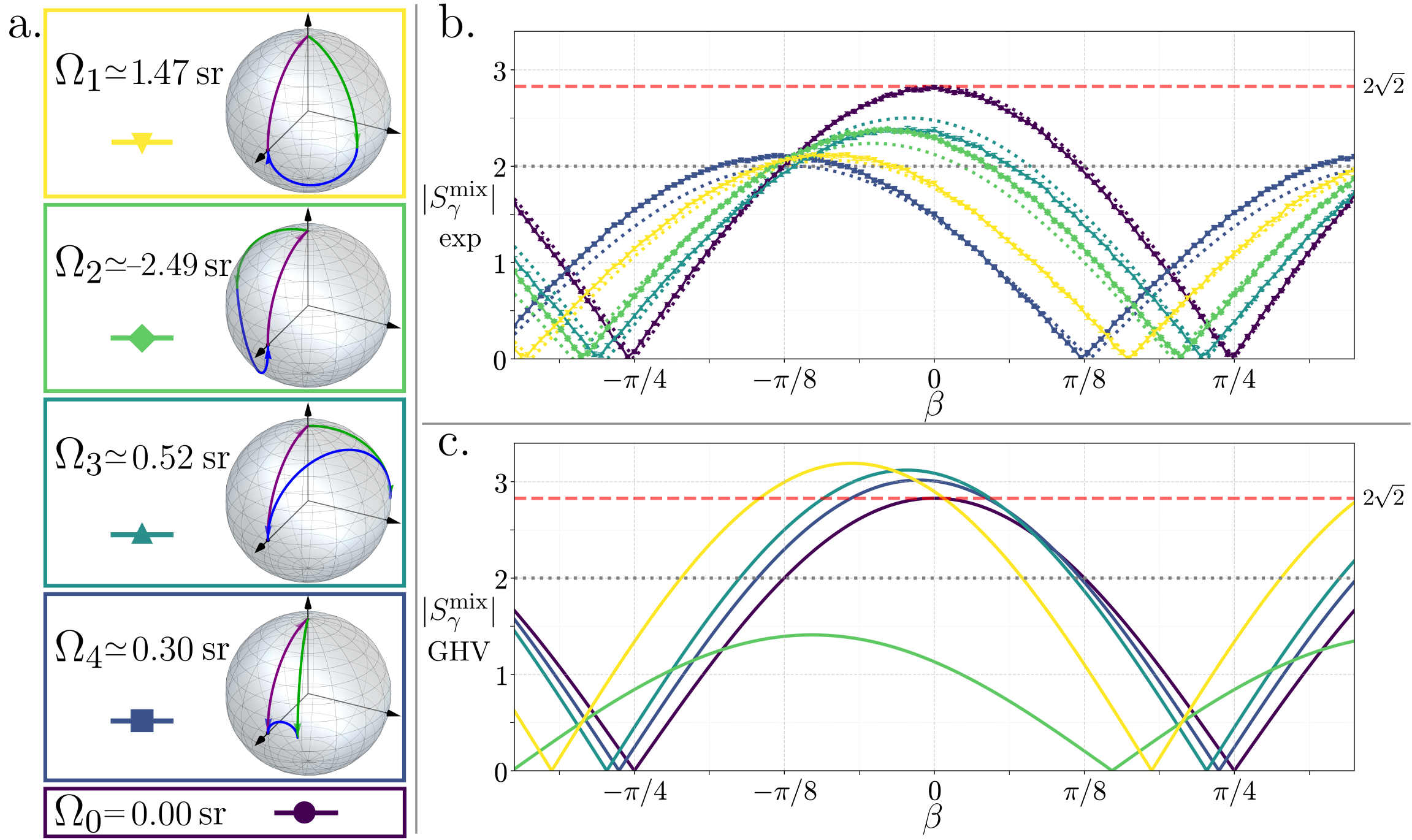}
    \caption{\textbf{Direct and looped implementations.} \textbf{a}, Closed trajectories tested on the Poincar\'e sphere; colors and symbols identify the loop parameter $\Omega_\gamma$. The loops are implemented using two QWPs and a HWP; their orientations are specified in the End Matter. \textbf{b}, Experimentally measured absolute mixed-context parameter $|S_\gamma^{\rm mix}|$ as a function of $\beta$. The dotted lines show the theoretical predictions of quantum mechanics -- the detailed derivations can be found in the End Matter. \textbf{c}, Eq.~\eqref{eq:ghv-mixed-fringe} for $V=1$ and $\delta_\gamma=\Omega_\gamma$. The grey dotted line at $2$ and the red dashed line at $2\sqrt2$ are shown as references to the standard CHSH and Tsirelson bounds, respectively; the contextual bound for the looped witness is instead given by Eq.~\eqref{eq:ghv-contextual-bound}. Error bars are smaller than the markers.}
    \label{fig:results}
\end{figure*}

The experimental witness mirrors the mixed-context construction in Eq.~\eqref{eq:ghv-contextual-chsh}. The first three correlations use the direct implementation, whereas the final occurrence of Bob's setting $b'$ is realized through the loop $\Gamma_\gamma$:
\begin{equation}
    {S}_\gamma^{\rm mix}(\beta)
    =E_0(a,b)+E_0(a,b')
    +E_0(a',b)-E_\gamma(a',b').
    \label{eq:ghv-experimental-mixed-chsh}
\end{equation}
For $a=0$, $a'=-\pi/4$, $b=\beta-\pi/8$ and $b'=\beta+\pi/8$, Eq.~\eqref{eq:ghv-loop-correlation} gives
\begin{equation}
    \begin{split}
    {S}_\gamma^{\rm mix}(\beta)
    =V\{&\cos(2\beta-\pi/4)
    +2\cos(2\beta+\pi/4)\\
    &-\cos(2\beta+3\pi/4+\delta_\gamma)\}.
    \end{split}
    \label{eq:ghv-mixed-fringe}
\end{equation}
The reference curve, obtained for $\delta_0=0$, is ${S}_0=2\sqrt2 V\cos(2\beta)$. Taking the signed difference $\Delta{S}_\gamma^{\rm mix}={S}_0-{S}_\gamma^{\rm mix}$ gives the particularly transparent relation
\begin{equation}
    \begin{split}
    \Delta{S}_\gamma^{\rm mix}(\beta)
    &=E_\gamma(a',b')-E_0(a',b')\\
    &=2V
    \sin\left(\frac{\delta_\gamma}{2}\right)
    \sin\left(2\beta
    -\frac{\pi}{4} +\frac{\delta_\gamma}{2}\right).
    \end{split}
    \label{eq:ghv-mixed-difference}
\end{equation}
For small $|\delta_\gamma|$, this reduces to $\Delta{S}_\gamma^{\rm mix}\simeq V\delta_\gamma\sin(2\beta-\pi/4)$. Because only one correlation term is changed, the predicted mixed-context curve is deformed in both position and amplitude; it is not a rigid translation of a standard CHSH fringe. The theoretical curves in Fig.~\ref{fig:results} use the illustrative choice $V=1$ and $\delta_\gamma=\Omega_\gamma$. They are predictions of the operational response in Eq.~\eqref{eq:ghv-loop-correlation}, not consequences of assigning a value to $p_\gamma$.\\

%%%%
\noindent\emph{Experimental implementation.}
Polarization-entangled photon pairs were generated by degenerate type-I spontaneous parametric down-conversion in two orthogonally oriented BiBO (BiB$_3$O$_6$) crystals. A quasi-continuous-wave laser at $355~\mathrm{nm}$ delivered an average power of approximately $100~\mathrm{mW}$. A half-wave plate before the crystals set the pump polarization and coherently excited the two down-conversion pathways. Their relative phase was compensated to produce a state close to $|\Phi^+\rangle$ with a fidelity of $96.4\%$. The experimental layout is shown in Fig.~\ref{fig:experiment}.

Alice and Bob analyzed their photons using a half-wave plate followed by a polarizing beam splitter. The outputs were coupled into single-mode fibers and detected with single-photon avalanche photodiodes. Interference filters centered near $710~\mathrm{nm}$ with 10~nm bandwidth selected the degenerate photons, and coincidences were recorded within a $2~\mathrm{ns}$ window. For each analyzer pair, the four coincidence counts were $N_{++}=N(\mathbf a,\mathbf b)$, $N_{-+}=N(\mathbf a_\perp,\mathbf b)$, $N_{+-}=N(\mathbf a,\mathbf b_\perp)$ and $N_{--}=N(\mathbf a_\perp,\mathbf b_\perp)$. The correlation was reconstructed as

\begin{equation}
    E(\mathbf a,\mathbf b)
    =\frac{N_{++}+N_{--}-N_{+-}-N_{-+}}
    {N_{++}+N_{--}+N_{+-}+N_{-+}}.
    \label{eq:experimental-correlation}
\end{equation}

Closed wave-plate sequences were then introduced in Bob's arm. Each sequence returned the analyzer to the same final measurement basis while tracing a path $\mathcal C_\gamma$ on the Poincar\'e sphere. Operational equivalence was established at the level of the final measurement effect before the entangled correlations were compared. To evaluate Eq.~\eqref{eq:ghv-experimental-mixed-chsh}, the correlations $E_0(a,b)$, $E_0(a,b')$, and $E_0(a',b)$ were measured with the direct analyzer, while $E_\gamma(a',b')$ was measured with the looped realization of the same nominal setting $b'$.

\noindent
\emph{Results.}
Figure~\ref{fig:results} compares the direct implementation with four closed trajectories in Bob's analyzer, labeled by $\Omega_1=\pi\left({4}/{3}-{\sqrt{3}}/{2}\right)\simeq+1.468$ sr, $\Omega_2=\pi\left({1}/{\sqrt{2}}-{3}/{2}\right)\simeq-2.491$ sr, $\Omega_3={\pi}/{6}\simeq+0.524$ sr and $\Omega_4=\pi\left(\sin({5\pi}/{9})-{8}/{9}\right)\simeq+0.301$ sr. Here $\Omega_\gamma$ labels the loop used in the wave-plate sequence; any identification with the hidden rotation $\delta_\gamma$ is a model assumption. In the illustrative theoretical curves we set $\delta_\gamma=\Omega_\gamma$, whereas the experiment tests whether the measured correlation retains any dependence on the loop at all.

The theoretical curves in Fig.~\ref{fig:results}b show the strong loop-dependent deformation expected if $\delta_\gamma=\Omega_\gamma$: changing the loop changes both the maximum and the position of $|S_\gamma^{\rm mix}|$. The measurements show a markedly different behavior. All five curves are in line with the predictions of quantum mechanics. The direct curve, $\Omega_\gamma=0$, is a standard CHSH test and violates the Bell-local bound. The looped quantities compare distinct complete implementations and should not be read as independent standard CHSH tests; their purpose is to test whether changing the implementation of $b'$ alters the observed correlation. No loop ordering of the kind predicted by the illustrative GHV response is observed.

The lower panel makes the null comparison directly. The signed residuals fluctuate about zero without the structured dependence on $\beta$ predicted by Eq.~\eqref{eq:ghv-mixed-difference}. Excursions in the displayed ratio become less informative near settings for which ${S}_{\rm ref}$ is small, because division by the reference amplifies both statistical noise and small calibration offsets. A quantitative constraint should therefore be obtained from the unnormalized signed difference, or directly from the four correlations, with visibility, offset and angular calibration treated as nuisance parameters. Such a fit determines each $\delta_\gamma$ and, once the enclosed areas $\Omega_\gamma$ are specified, the coupling $\kappa$. No loop-dependent phase is resolved in the present data.
% Report the fitted delta_gamma values, their 95% confidence intervals and the
% goodness of fit here once the final count-level analysis is available.

\noindent\emph{Discussion and conclusion.}
The local tangent-frame construction and the experimentally tested joint response play different roles. The former shows how a hidden variable can retain geometric path memory while leaving the local marginal unchanged. It also defines the coupling-dependent mismatch probability $p_\gamma$. Because the construction remains Bell factorisable, however, it cannot reproduce the reference Bell violation. Equation~\eqref{eq:ghv-contextual-bound} is consequently a resource bound on a mixed-context counterfactual expression, not a derivation of the quantum value.

The operational model in Eqs.~\eqref{eq:ghv-operational-joint} and \eqref{eq:ghv-loop-correlation} begins instead from the observed non-factorisable correlation and asks whether a hidden holonomy changes the looped term in the mixed-context witness. The data address this second question. They constrain $\delta_\gamma$ and, once the geometric areas are specified, the coupling $\kappa$. They do not determine $p_\gamma$ without an additional coupling that consistently embeds the non-factorisable response. In particular, a null $\Delta S_\gamma^{\rm mix}$ does not imply that all counterfactual hidden outcomes are unchanged.

A complete GHV account of the quantum Bell correlation would therefore require nonlocal or otherwise non-Bell-factorisable structure. One possible direction is to supplement the local fiber variables with a relational variable $\Lambda_{AB}$ connecting Alice's and Bob's frames. Such a theory would explicitly give up Bell locality while retaining operational no-signaling after averaging. The present experiment does not test an arbitrary relational construction: only Bob's implementation path is varied. A two-sided test, in which Alice's and Bob's trajectories are changed independently, would be required to isolate contributions depending on their joint geometry.

The experiment addresses a question distinct from Bell locality: whether a measurement is fully specified by its final quantum effect or whether its implementation history leaves an additional trace in entangled correlations. Within the present precision, calibrated paths leading to the same analyzer give the same mixed-context witness. The result agrees with quantum mechanics and places direct constraints on path-sensitive hidden-variable models that predict an observable memory of the optical trajectory.\\

\noindent\emph{Acknowledgments:}
The authors acknowledge fruitful discussions with Justin Dressel, Eliahu Cohen, Lukas Burns and Marko \'Cosi\'c. A large language model is used for language editing and proofreading. The authors take full responsibility for the scientific content and conclusions of the manuscript.

%\bibliographystyle{unsrt}
%\bibliography{bibliography}

\vspace{1cm}

\section*{End matter}
%\textbf{Quantum-mechanical description of polarization correlations}
We consider polarization-entangled photon pairs initially prepared in the Bell state $|\Phi^+\rangle=\frac{1}{\sqrt2}(|HH\rangle+|VV\rangle)$. In Bob's arm, two quarter-wave plates (QWPs) followed by a half-wave plate (HWP) are introduced, with optical axes oriented at \(\theta_1=\pi/4\), \(\theta_2=\pi/4+\omega\), and \(\theta_3=\pi/4+\omega/2\), respectively. The resulting Jones matrix is
\begin{align}
U(\omega)=-
\begin{pmatrix}
e^{-i\omega}&0\\
0&e^{i\omega}
\end{pmatrix}.
\end{align}

This transformation returns horizontal and vertical polarizations to themselves while introducing a relative phase \(2\omega\). Consequently, the initial Bell state becomes, up to a global phase,
\begin{align}
|\Phi_\omega\rangle=
\frac{1}{\sqrt2}
\left(|HH\rangle+e^{i2\omega}|VV\rangle\right).
\end{align}

For linear polarization measurements at angles $\alpha$ and $\beta$, the correlation function is
\begin{eqnarray}
E_\omega(\alpha,\beta)&=&\langle\Phi_\omega|M(\alpha)\otimes M(\beta)|\Phi_\omega\rangle\cr\nonumber
&=&\cos2\alpha\cos2\beta+\cos2\omega\sin2\alpha\sin2\beta,
\end{eqnarray}
where $M(\theta)=\cos2\theta\,\sigma_z+\sin2\theta\,\sigma_x$.

Using the experimental settings $a=0,\, a'=-\pi/4,\, b=\beta-\pi/8$, and $b'=\beta+\pi/8$, the mixed-context Bell parameter, with the waveplates applied only to Bob's fourth measurement, becomes
\begin{align}
S_{\omega,\mathrm{QM}}^{\mathrm{mix}}(\beta)=\frac{1}{\sqrt2}\left[(3+\cos2\omega)\cos2\beta+(\cos2\omega-1)\sin2\beta\right].
\end{align}

For the direct measurement, $S_0=2\sqrt2\cos2\beta$. The difference is therefore
\begin{align}
S_0-S_{\omega,\mathrm{QM}}^{\mathrm{mix}}
=(1-\cos2\omega)\sin(2\beta+\pi/4).
\end{align}

\end{document}